\documentclass[aps,prl,twocolumn,superscriptaddress,groupedaddress,preprintnumbers]{revtex4-2} 
\usepackage[totalheight = 23cm, totalwidth = 17cm]{geometry}
\usepackage{amssymb,amsmath,amsfonts,amsbsy}
\usepackage{comment}
\usepackage{graphicx}  
\usepackage{dcolumn}   
\usepackage{bm}        
\usepackage{amssymb}   
\usepackage{amsmath}
\usepackage{mathtools}
\usepackage{cases}
\usepackage{mathrsfs}
\usepackage{color}
\usepackage{xcolor}
\usepackage{enumitem}
\usepackage{ulem}

\makeatletter
\newcommand*\bigcdot{\mathpalette\bigcdot@{.5}}
\newcommand*\bigcdot@[2]{\mathbin{\vcenter{\hbox{\scalebox{#2}{$\m@th#1\bullet$}}}}}
\makeatother

\newcommand{\be}{\begin{equation}}
	\newcommand{\ee}{\end{equation}}
\newcommand{\bea}{\begin{eqnarray}}
	\newcommand{\eea}{\end{eqnarray}}

\newcommand{\dif}{\mathop{}\!\mathrm{d}}

\begin{document}
	\preprint{YITP-25-17}
	\title{$\delta n$ formalism: \\
		A new formulation for the probability density of the curvature perturbation}
	\author{Diego Cruces${}^{a}$}
	\email{dcruces@itp.ac.cn}
	\author{Shi Pi${}^{a,b,c}$} \email{shi.pi@itp.ac.cn} 
	\author{Misao Sasaki${}^{c,d,e,f}$}\email{misao.sasaki@ipmu.jp}
	\affiliation{
		$^{a}$ Institute of Theoretical Physics, Chinese Academy of Sciences, Beijing 100190, China \\
		$^{b}$ Center for High Energy Physics, Peking University, Beijing 100871, China\\
		$^{c}$ Kavli Institute for the Physics and Mathematics of the Universe (WPI), The University of Tokyo, Kashiwa, Chiba 277-8583, Japan\\
		$^{d}$ Asia Pacific Center for Theoretical Physics (APCTP), Pohang 37673, Korea \\
		$^{e}$ Center for Gravitational Physics and Quantum Information,
		Yukawa Institute for Theoretical Physics, Kyoto University, Kyoto 606-8502, Japan\\
		$^{f}$ Leung Center for Cosmology and Particle Astrophysics,\\
		National Taiwan University, Taipei 10617}
	\date{\today}
	\begin{abstract}
		$\delta N$ formalism is a useful method to calculate the curvature perturbation on superhorizon scales. Contrary to what it is typically done in the literature, we re-formulate the $\delta N$ formalism by using the $e$-folding number $n$ counted forward in time. For a fixed initial time $\bar{n}_0$, the probability density function (PDF) of the field perturbation $\delta\phi_0$ and its velocity $\delta\pi_0$ are specified by the solutions of the perturbation equation on subhorizon scales. As $\delta\pi_0$ is fully correlated with $\delta\phi_0$ soon after horizon exit, we find a novel $\delta n$ formalism to calculate the curvature perturbation as well as its PDF.
		It can naturally incorporate those trajectories for which inflation never ends, which are lost when counting $N$ backward in time. 
	\end{abstract}
	\maketitle

	\textit{Introduction.}--- 
	Due to the quantum nature of inflationary fluctuations \cite{Mukhanov:1985rz,Sasaki:1986hm}, there is always some probability of randomly generating a large enough fluctuation such that the overdensity generated at horizon reentry will be so large that it will collapse into a singularity, giving rise to the so-called Primordial Black Hole (PBH) \cite{Zeldovich:1967lct,Hawking:1971ei,Carr:1974nx,Meszaros:1974tb,Carr:1975qj,Khlopov:1985jw}. Since those overdensities originate from the primordial curvature perturbation on comoving slices $\mathcal{R}$ at the end of inflation (see \cite{Escriva:2022duf} for a review), it is of crucial importance to know the probability density function (PDF) of $\mathcal{R}$, especially its tail, to correctly determine the masses and abundances of PBH \cite{Young:2013oia,Young:2015cyn,Atal:2018neu,Passaglia:2018ixg,Yoo:2019pma,Kehagias:2019eil,Mahbub:2020row,Riccardi:2021rlf,Davies:2021loj,Young:2022phe,Escriva:2022pnz,Matsubara:2022nbr}.
	
	At linear order in cosmological perturbation theory, $\mathcal{R}$ is Gaussian and hence its PDF is fully determined by its 
	variance, which is given by the power spectrum $\mathcal{P}_{\mathcal{R}}$.
	The observed curvature perturbation at CMB scales is nearly Gaussian \cite{Planck:2018jri}, and hence compatible with linear perturbation theory. Furthermore, the power spectrum at these scales is of order $10^{-9}$, 
	implying an exponentially suppressed, completely negligible probability of generating a large ($\sim \mathcal{O}(0.1)$) fluctuation that eventually collapse into a PBH.
	
	However, $\mathcal{P}_{\mathcal{R}}$ can be much enhanced at smaller scales, where there is no observational constraint. In fact, many inflation models predict such a growth, especially the single-field inflation with a segment of flat or bumpy potential 
	which invokes non-attractor behavior
	\cite{Ivanov:1994pa, Bullock:1996at, Germani:2017bcs, Hertzberg:2017dkh, Garcia-Bellido:2017mdw, Pi:2022zxs, Pi:2022ysn, Wang:2024wxq}. The enhancement of $\mathcal{P}_{\mathcal{R}}$ has two important consequences: 1) Large fluctuations which eventually collapse to PBHs become more probable and 2) Non-linear effects alters the statistics of the fluctuation, especially deforms the tail of its PDF.
	Thus, the precise knowledge of the tail of the PDF for $\mathcal{R}$ motivate us to go beyond linear perturbation theory by employing methods such as spatial gradient expansion \cite{Nambu:1994hu, Sasaki:1998ug, Tanaka:2007gh}.
	
	The idea of gradient expansion is to define an expansion parameter $\sigma \equiv \frac{k}{a H} \ll 1$ as the ratio between the comoving Hubble radius $(a H)^{-1}$ and the characteristic scale of the fluctuation $L \sim k^{-1}$. At leading order one can show that each point evolves like a separate homogeneous universe with different values of Hubble rate, scale factor, inflaton field value, etc. Therefore, these values satisfy their PDFs among all the seperate universes, and this leading order spatial gradient expansion is usually dubbed as ``separate universe approach" (SUA) \cite{Wands:2000dp,Lyth:2004gb}. 
	
	In the $\delta N$ formalism, the SUA is combined with initial conditions on a spatially flat slice given by perturbation theory (typically linear) in each Hubble patch to compute the $e$-folding number from this initial slice to the comoving slice at a late epoch $t=t_{\rm f}$. 
	The difference of the number of $e$-folds $N$ in such a separate universe and the fiducial $\bar{N}$ gives the
	curvature perturbation at $t_{\rm f}$ in each patch, \textit{i.e.} $\mathcal{R}=\delta N=N-\bar{N}$ 
	\cite{Salopek:1990jq,Sasaki:1995aw,Sugiyama:2012tj, Abolhasani:2019cqw}. We will keep using the bar notation to denote quantities defined in the fiducial background throughout the whole paper. Typically, the $\delta N$ formalism is formulated by the difference of $N$ counted backward in time from a fixed final comoving hypersurface at the moment when inflation ends, 
	or after the universe has entered the last attractor regime.
	This allows us to define a fixed $e$-folding number $\bar{N}_\mathrm{f}$ (usually set to 0) at that epoch, from which we start counting $N$ backward in time in every local patch. 
	The difference in the total $e$-folding number $N$ for different patches originates from the difference in the initial conditions on the initial spatially-flat hypersurface. 
	See \cite{Abolhasani:2013zya} for a detailed discussion.

	Although very powerful, the $\delta N$ formalism has only been applied to quite simple inflationary models analytically, such as slow-roll \cite{Lyth:2005fi}, ultra-slow-roll \cite{Cai:2018dkf,Biagetti:2018pjj}, constant-roll \cite{Atal:2019erb,Atal:2019cdz}, or in general a (piecewise) quadratic potential \cite{Pi:2022ysn, Wang:2024wxq}. The reason is that, as we will see, although $\phi$ and $\pi(=-d\phi/dN)$ may be relatively easy to compute, expressing $N$ as a function of $(\phi,\pi)$, $N(\phi,\pi)$, which is required to compute the PDF of $\delta N$, is not easy.
	In particular, it is analytically formidable when there is some non-trivial dynamics of the inflaton field, and it is often a challenge even numerically.
	
	In this paper, we present a novel scheme that highly simplifies the computation of the PDF of $\delta N$, by using the $e$-folding number $n$ counted forward in time. To do so, we take some initial time $\bar{n}_0$ to be the same for all local patches described by SUA. Then we count the number of $e$-folds $n$ forward in time for each local patch until it reaches the final comoving hypersurface, 
	which is different for different patches. 
	
	\textit{The $\delta n$ formalism.}---
	In the 3+1 (or ADM) formulation of general relativity \cite{Arnowitt:1959ah}, the metric takes the form, 
	\begin{equation}
		\dif s^2=-\alpha^2 \dif t^2 +\gamma_{ij}\left(\dif x^i+\beta^i \dif t\right)\left(\dif  x^j+\beta^j\dif t\right)\,,
		\label{ADM_metric}
	\end{equation}
	where $\alpha$ is the lapse function, $\beta^i$ is the shift vector, and the spatial metric can be redefined as
	$\gamma_{ij}\equiv a(t)^2 e^{2 \psi}\tilde{\gamma}_{ij}$,
	with $\det \left(\tilde{\gamma}_{ij}\right)=1$. 
	We set $a(t)$ to be given by an isotropic and homogeneous background solution, which we regard as a fiducial background.
	
	We denote the $e$-folding number counted forward in time by $n$, and its differential is defined by
	\begin{equation}
		\dif n(t,\bm{x}) \equiv \frac{1}{3}K(t, \bm{x})\alpha(t, \bm{x})\dif t\,,
		\label{def_dN}
	\end{equation}
	where
	\begin{equation}
		K = 3\frac{H}{\alpha} + 3 \frac{\dot{\psi}}{\alpha} - \frac{D_i \beta^i}{\alpha}\,,
		\label{K}
	\end{equation}
	is the trace of the extrinsic curvature, with $H\equiv \dot a/a$ and $D_i$ the spatial covariant derivative. 
	The variable $\bm{x}$ represents the position $x^i$, which labels each Hubble patch. 
	In the language of gradient expansion, SUA is the limit $H^{-1}\to0$ while keeping the length of interest $L$ finite. 
	Integrating \eqref{def_dN}, we find that the number of $e$-folds counted forward in time from some background  initial time $\bar{n}_{0}$, which is set to be the same for all local patches, to some final time $n_\mathrm{f}$,
	\begin{equation}
		n_\mathrm{f}(t_0,t_\mathrm{f},\bm{x})-\bar{n}_0 =  \frac{1}{3}\int_{t_0}^{t_\mathrm{f}}K(t', \bm{x})\alpha(t', \bm{x}) dt'\,.
		\label{def_N}
	\end{equation}
	The SUA guarantees that the last term in \eqref{K} decays as the inverse volume \cite{Cruces:2022dom}, thus \eqref{def_N} is simplified to
	\begin{equation}
		n_\mathrm{f}\left(t_0, t_\mathrm{f}, \bm{x}\right) = \bar{n}_\mathrm{f}\left(t_0,t_\mathrm{f}\right) + \psi(t_\mathrm{f}, \bm{x}) - \psi(t_0, \bm{x})\,,
		\label{N_gradient}
	\end{equation}
	where 
	$\bar{n}_\mathrm{f}\left(t_0,t_\mathrm{f}\right)-\bar{n}_{0}\equiv\int_{t_0}^{t_\mathrm{f}}H\left(t'\right)dt'$ is the background number of $e$-folds.
	
	We can finally define $\delta n$ as
	\begin{equation}
		\delta n
		\left(t_0, t_\mathrm{f},\bm{x}\right) \equiv \left(n_\mathrm{f}\right)_{\Sigma_{\rm i}}^{\Sigma_{\rm f}}\left(t_0, t_\mathrm{f}, \bm{x}\right) - \bar{n}_\mathrm{f}\left(t_0,t_\mathrm{f}\right)\,,
		\label{deltaN_definition}
	\end{equation}
	where the subscript $\Sigma_{\rm i}$ and superscript $\Sigma_{\rm f}$ are to emphasize the difference in the time slicing conditions at the initial time $t_0$ and final time $t_\mathrm{f}$, respectively. 
	In the $\delta N$ (as well as in the $\delta n$)formalism,  
	we choose an initial hypersurface to be spatially-flat $\psi(t_0,\bm{x})=0$ and a final hypersurface to be comoving $\psi(t_\mathrm{f},\bm{x})=\mathcal{R}$, so that \eqref{N_gradient} becomes
	\begin{equation}
		\delta n \left(t_0,t_\mathrm{f},\bm{x}\right) = \mathcal{R}\left(t_\mathrm{f},\bm{x}\right)\,,
		\label{deltaN_curvature}
	\end{equation}
	where $\mathcal{R}$ is the non-linear comoving curvature perturbation of the patch around $\bm{x}$ on the final hypersurface. This coincides exactly with the conventional $\delta N$ formula.
	
	Once we have identified $\delta n$ with 
	the curvature perturbation,  
	the only task left is to compute $\delta n$ itself. 
	To clearly demonstrate the power of the new formalism, let us focus on single-field inflation  with a canonical kinetic term, of which the action is
	\begin{equation}
		S = \int d^4x \sqrt{-g}\left[\frac{M_\mathrm{Pl}^2}{2}R-\frac12\left(\partial\phi\right)^2-V(\phi)\right],
		\label{action}
	\end{equation}
	where $R$ is the Ricci scalar of the metric $g_{\mu\nu}$. 
	The equations of motion that describe the time evolution of each local patch 
	in the context of SUA are easily obtained from \eqref{action},
	\begin{equation}\label{EOM_patch}
		\begin{aligned}
			\frac{\partial \phi}{\partial n} & = \pi\,, \\
			\frac{\partial \pi}{\partial n} & + \left(3-\frac{\pi^2}{2M_\mathrm{Pl}^2}\right)\left(\pi + \frac{M_\mathrm{Pl}^2}{V(\phi)}\frac{d V(\phi)}{d \phi}\right)=0\,.
		\end{aligned}
	\end{equation}
	The solution is given by
	\begin{equation}\label{sol_EOM_patch}
		\begin{aligned} 
			\phi_0\left(n_\mathrm{f}(\bm{x})-\bar{n}_{0},\bar{\phi}_\mathrm{f},\pi_\mathrm{f}(\bm{x})\right)\,, \\ \pi_0\left(n_\mathrm{f}(\bm{x})-\bar{n}_{0},\bar{\phi}_\mathrm{f},\pi_\mathrm{f}(\bm{x})\right)\,,
		\end{aligned}
	\end{equation}
	where we have used the fact that the final hypersurface is comoving, $\phi_\mathrm{f}(\bm{x})=\bar{\phi}_\mathrm{f}$. 
	We can invert \eqref{sol_EOM_patch} to get $\pi_\mathrm{f}(\phi_0,\pi_0)$ and $n_\mathrm{f}(\phi_0,\pi_0)$. Here and in what follows, the dependence of $\phi_0$ and $\pi_0$ on $\bm{x}$ is suppressed, with the understanding that any quantity without a bar is automatically $\bm{x}$-dependent. 
	The variation of $\phi_0$ and $\pi_0$ in the field and momentum space (which originates from the spatial variation thanks to SUA) gives rise to the variation of the number of $e$-folds, 
	\begin{align}
		\delta n =  n_\mathrm{f}\left(\bar{\phi}_0+\delta\phi_0,\bar{\pi}_0+\delta\pi_0\right)-\bar{n}_\mathrm{f}\left(\bar{\phi}_0,\bar{\pi}_0\right),
		\label{deltaN_back}
	\end{align} 
	where the fluctuations $\delta\phi_0$ and $\delta\pi_0$ are those on spatially flat slices by definition.	
	
\begin{figure}[h]
		\centering
		\includegraphics[width=0.5\textwidth]{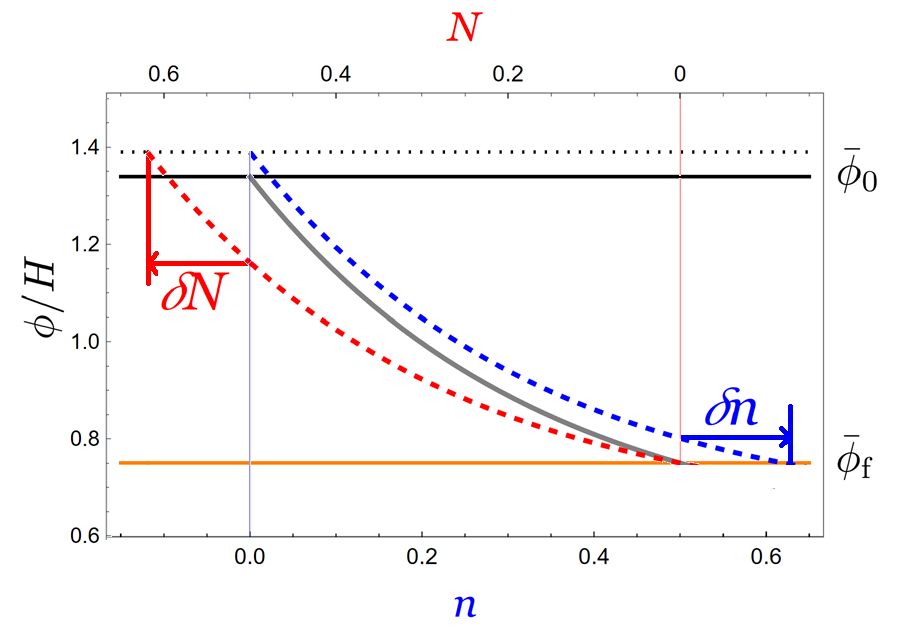}
		\caption{Schematic picture showing the equivalence of $\delta N$ (in red) and $\delta n$ (in blue). 
			The red and blue dotted lines represent the perturbed initial conditions $\bar{\phi}_0+\delta\phi_0$ with fixed number of $e$-folds at the end of inflation and at the initial time, respectively.
			We also show with a solid gray line the fiducial background trajectory for better visualization.
		}	\label{fig:equivalence_USR}
	\end{figure}

	The above discussion is parallel to the ordinary $\delta N$ formalism, in which $\bar{N}_\mathrm{f}$ is fixed and the $e$-folding number is counted backward in time. See \text{e.g.} \cite{Abolhasani:2019cqw}. 
	The results are equivalent, showing the exact equality of $\mathcal{R}=\delta N=\delta n$. 
	A schematic comparison of these two methods is shown in Fig. \ref{fig:equivalence_USR}.
	
	Finally, once we know the function $\delta n(\delta\phi_0,\delta\pi_0)$, we can easily write the PDF of $\delta n$, \textit{i.e.} $\mathcal{R}$, by using the conservation of probability,
	\begin{align}	
		\label{PDF_deltaN_generic}
		&\mathbb{P}_{\delta n}(\delta n)=\iint\dif\delta\phi_0\dif\delta\pi_0
		\\
		\nonumber
		&\quad\times\mathbb{P}_{\delta\phi,\delta\pi}(\delta\phi_0, \delta\pi_0)\,\boldsymbol{\delta}\Big(\delta n-\delta n(\delta\phi_0,\delta\pi_0)\Big)\,,	
	\end{align}
	where $\boldsymbol{\delta}$ is the Dirac delta function.
	
	From \eqref{PDF_deltaN_generic}, we can see that the PDF of the non-linear curvature perturbation $\mathcal{R}$ obviously depends on that of $\delta\phi_0$ and $\delta\pi_0$ on the initial hypersurface, which are usually Gaussian as they originate from quantum fluctuations and obey the linear
	perturbation equation.
	The intrinsic non-Gaussianities of $\mathbb{P}_{\delta\phi,\delta\pi}(\delta\phi_0, \delta\pi_0)$ from self-interactions can be included straightforwardly as in Ref.~\cite{Ballesteros:2024pbe}.
	
	Throughout the following discussion, the mode of interest we consider is 
	\begin{equation}
		k_{\sigma}=\sigma a(\bar{n}_0) \bar{H}(\bar{n}_0)\,;
		\quad\sigma \ll1\,,
	\end{equation}
	so that we are on superhorizon scales at time $\bar{n}_0$, ensuring that the SUA (and hence the $\delta n$ formalism) can be consistently applied. 
	
	At linear order in perturbation theory, the PDF $\mathbb{P}_{\delta\phi,\delta\pi}(\delta\phi_0, \delta\pi_0)$ for the mode $k_{\sigma}$ is Gaussian with zero mean, 
	\begin{equation}\label{PDF_full}
		\frac1{2\pi\sqrt{\mathrm{det}\boldsymbol{\Sigma}}}
		\exp\left[-\frac12(\delta\phi_0,\;\;\delta\pi_0)\boldsymbol{\Sigma}^{-1}\left(
		\begin{matrix}
			\delta\phi_0\\
			\delta\pi_0
		\end{matrix}
		\right)\right],
	\end{equation}
	with the covariance matrix $\Sigma$ given by \cite{Grain:2017dqa, Agullo:2022ttg}
	\begin{align} \nonumber
		\boldsymbol{\Sigma} &\equiv 
		\left(\begin{matrix}
			\sigma^2_{\delta\phi,\delta\phi}(k_{\sigma},\bar{n}_0) & \sigma^2_{\delta\phi,\delta\pi}(k_{\sigma},\bar{n}_0) \\
			\\	
			\sigma^2_{\delta\pi,\delta\phi}(k_{\sigma},\bar{n}_0) & \sigma^2_{\delta\pi,\delta\pi}(k_{\sigma},\bar{n}_0)
		\end{matrix} 
		\right),
	\end{align}
	where 
	\begin{align}
		&	\sigma^2_{\delta X,\delta Y}(k,n)\equiv \text{Re}\left[\frac{k^3}{2\pi^2}\delta X_k(n)\delta Y_k^{\star}(n)\right]\,;
		\nonumber\\
		&\quad X,\,Y=\phi~ {\rm or}~\pi\,,
		\label{variance_def}
	\end{align}
	and the superscript $\star$ denotes the complex conjugate, $\delta\phi_{k_{\sigma}}(\bar{n}_0)$ and $\delta\pi_{k_{\sigma}}(\bar{n}_0)$ are those obtained by solving the perturbation equations, 
	\begin{align} \label{MS_equation}
		&\delta\pi_k = \frac{\partial \delta\phi_k}{\partial n} \,, \\ \nonumber
		&0=\frac{\partial \delta\pi_k}{\partial n}  + \left(3-\epsilon_1\right)\delta\pi_k+\\\nonumber
		&\left[\left(\frac{k}{aH}\right)^2+\left(-\frac{3}{2}\epsilon_2+\frac{1}{2}\epsilon_1\epsilon_2-\frac{1}{4}\epsilon_2^2-\frac{1}{2}\epsilon_2\epsilon_3\right)\right]\delta\phi_k\,,
	\end{align}
	and $\epsilon_1=-\dot{H}/ H^2$, 
	$\epsilon_2=-\dot\epsilon_{1}/(H\epsilon_{1})$, 
	$\epsilon_3=-\dot\epsilon_{2}/(H\epsilon_{2})$ 
	are the slow-roll parameters.
	
	It can be shown \cite{Figueroa:2021zah,Chandran:2023ogt,Mishra:2023lhe,Noorbala:2024fim} that the solution on super-horizon scales ($\sigma\ll1$) 
	satisfies 
	\begin{equation} 
		\frac{\text{Im}\left[\delta\phi_{k_{\sigma}}(\bar{n}_0)\delta\pi_{k_{\sigma}}^{\star}(\bar{n}_0)\right]}{\text{Re}\left[\delta\phi_{k_{\sigma}}(\bar{n}_0)\delta\pi_{k_{\sigma}}^{\star}(\bar{n}_0)\right]}\ll 1\,,
		\label{determinant}
	\end{equation}
	which implies
	\begin{equation}
		\frac{\det \left[\boldsymbol{\Sigma}\right]}{\sigma^2_{\delta\phi,\delta\phi}\sigma^2_{\delta\pi,\delta\pi}}\ll 1\,.
		\label{cond_determinant}
	\end{equation}
	This inequality \eqref{cond_determinant} is important
	because it means that $\delta\phi_0$ and $\delta\pi_0$ are fully correlated on superhorizon scales. 
	In other words, 
	\begin{equation} 
		\rho\equiv \frac{\sigma^2_{\delta\pi,\delta\phi}(k_{\sigma},\bar{n}_0)}{\sigma_{\delta\phi,\delta\phi}(k_{\sigma},\bar{n}_0) \sigma_{\delta\pi,\delta\pi}(k_{\sigma},\bar{n}_0)}\simeq\pm 1\,,
		\label{pearson}
	\end{equation}
	which gives $\rho=\text{sign}\left(\text{Re}\left[\delta\phi_{k_{\sigma}}(\bar{n}_0)\delta\pi_{k_{\sigma}}^{\star}(\bar{n}_0)\right]\right)$, or equivalently,
	\begin{equation}
		\delta\pi_0=\pm \frac{\sigma_{\delta\pi,\delta\pi}(k_{\sigma},\bar{n}_0)}{\sigma_{\delta\phi,\delta\phi}(k_{\sigma},\bar{n}_0)}\delta\phi_0\,.
		\label{correlation_deltapi_deltaphi}
	\end{equation}
	Setting $|\rho|=1$ in \eqref{PDF_full}, 
	$\mathbb{P}_{\delta\phi,\delta\pi}(\delta\phi_0, \delta\pi_0)$ reduces to
	\begin{align} \label{PDF_deltaphi_deltapi}
		&\quad\;\mathbb{P}_{\delta\phi,\delta\pi}(\delta\phi_0, \delta\pi_0)\\
		&=\boldsymbol{\delta}\left(\frac{\delta\pi_0}{\sigma_{\delta\pi,\delta\pi}} \mp \frac{\delta\phi_0}{\sigma_{\delta\phi,\delta\phi}}\right)
		\frac{\exp\left[-\frac18\left(\frac{\delta\pi_0}{\sigma_{\delta\pi,\delta\pi}} \mp \frac{\delta\phi_0}{\sigma_{\delta\phi,\delta\phi}}\right)^2\right]}
		{\sqrt{2\pi}\sigma_{\delta\phi,\delta\phi}\sigma_{\delta\pi,\delta\pi}}\,.
		\nonumber
	\end{align}
	It then immediately follows that $\mathbb{P}_{\delta n}(\delta n)$ in \eqref{PDF_deltaN_generic} is given by
	\begin{equation}
		\mathbb{P}_{\delta n}(\delta n)=\mathbb{P}_{\delta\phi}(\delta\phi_0 (\delta n))\left| \frac{d\delta\phi_0 (\delta n)}{d\delta n}\right|\,. \label{PDF_onlydeltaphi}
	\end{equation}
	It is important to note that this formula implies we only need to know $\delta\phi_0$ as a function of $\delta n$, but not $\delta n$ as a function of $\delta\phi_0$.
	Note also that \eqref{PDF_onlydeltaphi} is valid only for $\sigma\ll1$, which is also the prerequisite of the SUA itself.
	
	\textit{A simplified method to compute $\mathbb{P}_{\delta n}$.}--- 
	Below we describe the procedure to compute $\delta\phi_0 (\delta n)$, which exhibits clear simplification in comparison with the usual $\delta N$ method that requires the 
	tedious expression of $\delta n (\delta \phi_0)$.
	
\begin{figure}[h]
		\centering
		\includegraphics[width=0.5\textwidth]{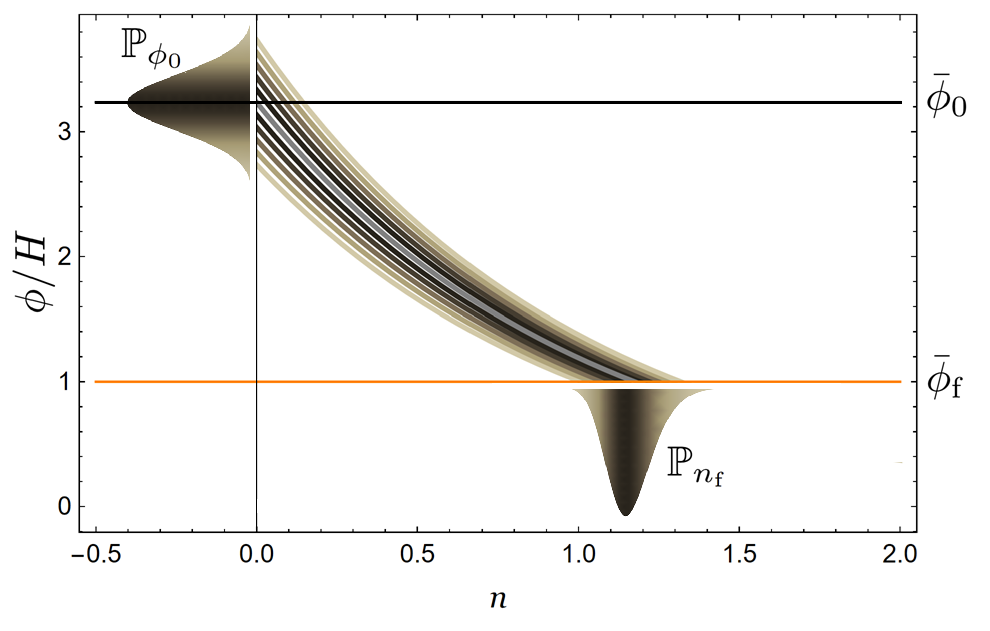}
		\caption{The procedure to obtain the PDF of $\delta n$, $\mathbb{P}_{n_{\rm f}}(\delta n)$ at the end of inflation from the initial PDF of $\phi_0$, $\mathbb{P}_{\phi_0}(\phi_0)$, in the example of constant-roll inflation with $m^2=(5/4)H^2$.
			Initially, $\mathbb{P}_{\phi_0}(\phi_0)$ is Gaussian with the mean $\bar{\phi}(\bar{n}_0)$ (solid black line) and the variance $\sigma^2_{\delta\phi, \delta\phi}(\kappa_{\sigma},\bar{n}_0)$, while $\pi_0$ is fully correlated with $\phi_0$, given by \eqref{correlation_deltapi_deltaphi}.
			The gray line represents the background trajectory $\bar{\phi}$, while the beige-black solid lines represent SUA trajectories where darker color means more probable trajectories. All trajectories terminate when the value of $\phi$ reaches $\phi_{\rm f}$ (orange solid line).
		}
		\label{fig:deltaN_simple}
	\end{figure}

	When counting $n$ forward in time, each SUA trajectory solves \eqref{EOM_patch} and it can be written as 
	\begin{equation}
		\phi\left(n-\bar{n}_0,\bar{\phi}_0+\delta\phi_0,\bar{\pi}_0+\delta\pi_0\right)\,,
	\end{equation}
	where the initial conditions are given at $n=\bar{n}_0$. The comoving slice condition for the final hypersurface  implies that there are no field fluctuations therein and hence,
	\begin{equation}
		\phi\left(n_\mathrm{f}-\bar{n}_0,\bar{\phi}_0+\delta\phi_0,\bar{\pi}_0+\delta\pi_0\right)=\bar{\phi}_{\mathrm{f}}\,.
		\label{gauge_condition_1}
	\end{equation}
	As it can be seen in Fig. \ref{fig:deltaN_simple}, the time $n_{\mathrm{f}}\equiv\bar{n}_{\mathrm{f}}+\delta n$ is obviously different for different trajectories. 
	Using that $\bar{\phi}_{\mathrm{f}}=\bar\phi(\bar n_{\rm f})$, 
	we can write \eqref{gauge_condition_1} as 
	\begin{align}
		\nonumber &\phi\left(\bar{n}_\mathrm{f}+\delta n-\bar{n}_0,\bar{\phi}_0+\delta\phi_0,\bar{\pi}_0\pm \frac{\sigma_{\delta\pi,\delta\pi}(k_{\sigma},\bar{n}_0)}{\sigma_{\delta\phi,\delta\phi}(k_{\sigma},\bar{n}_0)}\delta\phi_0\right)\\&=\phi\left(\bar{n}_\mathrm{f}-\bar{n}_0,\bar{\phi}_0,\bar{\pi}_0\right)\,,   \label{gauge_condition_2}
	\end{align}
	where we have used \eqref{correlation_deltapi_deltaphi} to write $\delta\pi_0$ in terms of $\delta\phi_0$. 
	
	Solving \eqref{gauge_condition_2} for $\delta n$
	is a must in the standard $\delta N$ formalism, 
	which is generically very difficult $\delta\phi_0$ \cite{Pi:2022ysn}. In contrast, solving \eqref{gauge_condition_2} for $\delta\phi_0$ is much simpler, which immediately leads to $\mathbb{P}_{\delta n}(\delta n)$ via \eqref{PDF_onlydeltaphi}. 
	This is the main advantage of our new method.
	
	As an example to show the power of this new approach, let us compute $\mathbb{P}_{\delta n}$ in a constant-roll model where $\delta n(\delta\phi_0)$ cannot be expressed analytically. 
	In the field range of our interest, the inflaton potential is quadratic, \textit{i.e.} $V(\phi)=V_0+\frac12m^2\phi^2$, of which the second slow-roll parameter is $\eta_V\equiv \frac{m^2}{3H^2}$, and the equations of motion are
	\begin{equation}
		\begin{aligned}
			\frac{\partial \phi}{\partial n}&=\pi\,,\\
			\frac{\partial \pi}{\partial n}&= -3\pi -3\eta \phi\,.
		\end{aligned}
	\end{equation}
	The general solution for $\phi$ can be written as
	\begin{align}
		\phi(n)&=c^{\,0}_{+}e^{-\lambda_{+}\left(n-\bar{n}_0\right)}+c^{\,0}_{-}e^{-\lambda_{-}\left(n-\bar{n}_0\right)}\,,
		\label{solution_quadratic_full}
	\end{align}
	where $\lambda_{\pm}\equiv \frac{3\pm\sqrt{9-12\eta_V}}{2}$ and $c^{\,0}_{\pm}\equiv \mp \frac{\pi_0+\lambda_{\mp}\phi_0}{\lambda_{+}-\lambda_{-}}$. Note that the solution proportional to
	$\exp\big(-\lambda_-(n-\bar{n}_0)\big)$ 
	is the attractor solution as it dominates when $n\gg1$.

	By solving the perturbation equation \eqref{MS_equation} and taking the long-wavelength limit, we obtain $\sigma_{\delta\phi,\delta\phi}(k,n)$,
	\begin{align}     \label{var_phi]}
		&\sigma^2_{\delta\phi,\delta\phi}(k,n)\\
		&= \left|C^{\,0}_{+}(k)e^{-\lambda_{+}(n-\bar{n}_0)}+C^{\,0}_{-}(k) e^{-\lambda_{-}(n-\bar{n}_0)}\right|^2\,,
		\nonumber
	\end{align}
	where $C^{\,0}_{\pm}(k)$ are constant coefficients determined either by the adiabatic vacuum condition or by matching conditions from previous stages (see e.g., \cite{Byrnes:2018txb,Pi:2022zxs,Wang:2024wxq,Briaud:2025hra}), depending on the model. Note that, although $C_{+}^{\,\mathrm{f}}$ is $k$-suppressed relative to $C_{+}^{\,\mathrm{f}}$
	under adiabatic vacuum conditions, and its inclusion would spoil condition \eqref{determinant}, this is no longer generically true when the initial conditions are set by matching conditions from previous stages. For this reason, and in order to be as general as possible, we include both modes in our analysis.
	
	We can similarly compute $\sigma_{\delta\pi,\delta\pi}(k,n)$ and solve \eqref{gauge_condition_2} to obtain
	\begin{align} \frac{\delta\phi_0}{\sigma_{\delta\phi,\delta\phi}(k_{\sigma},\bar{n}_0)}=
		\frac{\bar{c}^{\,\mathrm{f}}_{+}\left(1-e^{-\lambda_{+}\delta n}\right)+\bar{c}^{\,\mathrm{f}}_{-}\left(1-e^{-\lambda_{-}\delta n}\right)}{\left|C^{\,\mathrm{f}}_{+}(k_{\sigma}) e^{-\lambda_{+}\delta n}+C^{\,\mathrm{f}}_{-}(k_{\sigma})e^{-\lambda_{-}\delta n }\right|},
		\label{h_quadratic}
	\end{align}
	where $\bar{c}^{\mathrm{\,f}}_{\pm}\equiv \mp \frac{\bar{\pi}_{\mathrm{f}}+\lambda_{\mp}\bar{\phi}_\mathrm{f}}{\lambda_{+}-\lambda_{-}}=\mp\frac{\bar{\pi}_{0}+\lambda_{\mp}\bar{\phi}_0}{\lambda_{+}-\lambda_{-}}e^{-\lambda{\pm}(\bar{n}_{\mathrm{f}}-\bar{n}_0)}$ and $C^{\mathrm{\,f}}_{\pm}(k_{\sigma})\equiv C^{\,0}_{\pm}(k_{\sigma})e^{-\lambda{\pm}(\bar n_{\rm f}-\bar n_0)}$.
	Finally,
	assuming the initial Gaussian PDF for the field, \eqref{PDF_onlydeltaphi} gives
	\begin{equation}
		\mathbb{P}_{\delta n}(\delta n)=\frac{1}{\sqrt{2\pi}}e^{-{\frac{\delta\phi_0^2}{2\sigma^2_{\delta\phi,\delta\phi}(k_{\sigma},\bar{n}_0)}}}\left|\frac{d}{d\delta n}\left(\frac{\delta\phi_0}{\sigma_{\delta\phi,\delta\phi}(k_{\sigma},\bar{n}_0)}\right)\right|,
		\label{PDF_quadratic}
	\end{equation}
	where $\frac{\delta\phi_0}{\sigma_{\delta\phi,\delta\phi}(k_{\sigma},\bar{n}_0)}$ is given by \eqref{h_quadratic}.
	
	We mention that \eqref{PDF_quadratic} reduces to some known PDFs obtained by the standard method when $\delta n(\phi_0,\pi_0)$ can be analytically obtained. 
	For example, assuming that the system has already reached the final attractor regime, we have
	\begin{align} \nonumber
		\bar{c}^{\,\mathrm{f}}_{+}=0\,,  \quad  &\bar{c}^{\,\mathrm{f}}_{-}=\bar{\phi}_{\mathrm{f}}\,,\\
		C^{\,\mathrm{f}}_{+}(k)=0\,,  \quad &C^{\,\mathrm{f}}_{-}(k)=\frac{\bar{H}}{2\pi}\frac{\Gamma\left[\frac{3}{2}-\lambda_{-}\right]}{\Gamma[\frac{3}{2}]}\left(\frac{k}{a_{\mathrm{f}}H}\right)^{\lambda_{-}}\,,
	\end{align}
	where $\Gamma$ is the Gamma function. In this case, we recover the well-known PDF \cite{Pi:2022ysn},
	\begin{equation}
		\mathbb{P}_{\delta n}(\delta n) = \frac{e^{\lambda_{-}\delta n}}{\sqrt{2\pi}\left|\frac{C^{\,\mathrm{f}}_{-}\left(k_{\sigma}\right)}{\bar{\pi}_\mathrm{f}}\right|} \exp\left[- \frac
		{\left(e^{\lambda_{-}\delta n}-1\right)^2}{2\lambda_{-}^2\left(\frac{C^{\,\mathrm{f}}_{-}\left(k_{\sigma}\right)}{\bar{\pi}_\mathrm{f}}\right)^2}\right]\,,
	\end{equation}
	where we have used $\bar{\pi}_{\mathrm{f}}=-\lambda_{-}\bar{\phi}_{\mathrm{f}}$.
	
	Another interesting example is a pure phase of ultra-slow-roll (USR) inflation ($\eta_V=0$). In this case we have 
	\begin{align} \nonumber
		\bar{c}^{\,\mathrm{f}}_{+}=-\frac{\bar{\pi}_{\mathrm{f}}}{3}\,, \quad  \bar{c}^{\,\mathrm{f}}_{-}=\frac{\bar{\pi}_{\mathrm{f}}}{3}+\bar{\phi}_{\mathrm{f}}\,,\\
		C^{\,\mathrm{f}}_{+}(k)=0\,, \quad C^{\,\mathrm{f}}_{-}(k)=\frac{\bar{H}}{2\pi}\,.
	\end{align}
	Therefore we immediately obtain the PDF at the end of the USR phase,
	\begin{equation}
		\mathbb{P}_{\delta n}(\delta n) = \frac{e^{-3\delta n}}{\sqrt{2\pi}\left|\frac{\bar{H}}{2\pi\bar{\pi}_\mathrm{f}}\right|} \exp\left[- \frac
		{\left(1-e^{-3\delta n}\right)^2}{18\left(\frac{\bar{H}}{2\pi\bar{\pi}_\mathrm{f}}\right)^2}\right]\,.
		\label{PDF_deltaN_USR_new}
	\end{equation}
	This coincides with the final PDF if the transition to a later slow-roll stage is abrupt \cite{Cai:2018dkf,Passaglia:2018ixg,Pi:2022ysn,Pi:2024jwt},  and clearly reproduces the exponential tail PDF \cite{Biagetti:2021eep,Gow:2022jfb,Pi:2022ysn}.
	
	\textit{Conclusion.}--- We reformulated the $\delta N$ formalism by counting the number of $e$-folds forward in time.
	Using the fact that on superhorizon scales the field and velocity fluctuations are completely correlated, $\delta\pi=\delta\pi(\delta\phi)$, we presented a simple and useful scheme to compute the PDF of the curvature perturbation $\mathbb{P}_{\delta n}$ at the end of inflation, which is crucial in calculating the PBH abundance.
	
	The new method is based on the fact that inflation ends on a comoving slice on which the value of the inflaton is homogeneous. Given the initial condition on a flat slice, this leads to \eqref{gauge_condition_2}, which is the condition that determines the nonlinear gauge transformation from the flat slice to the comoving slice. Solving this equation for $\delta\phi_0$ as a function of $\delta n$ is far simpler than solving it for $\delta n$ as a function of $\delta\phi_0$. 
	
	The advantages of the new \textit{$\delta n$ formalism} are two-fold. One is its practicality. As illustrated in Fig.\ref{fig:deltaN_simple}, the numerical implementation of this method is very simple: One just has to numerically solve the background equations of motion with different perturbed initial conditions until each trajectory reaches $\phi_{\mathrm{f}}$. Then giving the weight to each of those trajectories according to the initial PDF of $\delta \phi_0$ immediately gives the final PDF of $\delta n$.
	
	Another is that our $\delta n$ formalism can help to solve the problem of probability loss. 
	Counting the number of e-folds $N$ along the trajectories backward in time from $\phi_\mathrm{f}$ can only cover a subset of the initial patches that can reach $\phi_\mathrm{f}$. 
	However, there are initial patches whose trajectories may stop evolving before reaching $\phi_\mathrm{f}$ due to an insufficient initial momentum.
	These patches contribute to $\mathbb{P}_{\phi_0}(\phi_0)$, but not to $\mathbb{P}_{n_\mathrm{f}}(n_\mathrm{f})$. They form false vacuum bubbles as the surrounding patches continue to roll down the potential, and finally collapse into Type B PBHs with an abundance comparable to ordinary PBHs from the adiabatic channel \cite{Escriva:2023uko,Escriva:2025ftp}. 
	With the new $\delta n$ formalism, such patches can be appropriately taken into account. 
	
	Furthermore, the novel $\delta n$ formalism may pave the way for a deeper understanding of the relationship between the classical and stochastic approaches \cite{Fujita:2013cna,Ballesteros:2024pwn, Jackson:2024aoo}. The advantage comes from the fact that, like the stochastic $\delta N$ formalism, our $\delta n$ formalism evolves perturbations forward in time. Importantly, in the limiting case of a single stochastic kick at $n=\bar{n}_0$ (where only one Fourier mode that crosses the horizon at $\bar n_0$ contributes), our method of computing the PDF of $\mathcal{R}$ is exactly equivalent to the first passage time analysis typically used in the stochastic $\delta N$ formalism \cite{Ivanov:1997ia,Vennin:2015hra}. 
	This connection supports the idea that the correlators computed by the stochastic $\delta N$ formalism are closely related with the Fourier transform of those computed by the classical $\delta N$ formalism, as recently suggested in \cite{Cruces:2024pni}.
	
	Although very powerful, this method relies on the validity of the SUA, i.e., the leading order approximation in spatial gradient expansion. 
	The SUA fails if the scale is not sufficiently far out from the Hubble horizon \cite{Jackson:2023obv}. In this case we must take into account higher-order terms in gradient expansion, and the $\delta N$ formalism must be formulated accordingly, as was done in \cite{Artigas:2024ajh}. Higher-order corrections in gradient expansion also break the complete correlation between $\delta\phi_k$ and $\delta\pi_k$. We leave the problem of finding a simple formula for the case of an incomplete correlation for future research.

	\textit{Acknowledgement.}---We thank Jaume Garriga for useful discussions. This work is supported by the National Key Research and Development Program of China Grant No. 2021YFC2203004, and
	JSPS KAKENHI grant No.~24K00624. 
	D.C. and S.P. are also supported by the National Natural Science Foundation of China (NSFC) Grants Nos. 12475066, 12447160 and 12447101. 
	
	\bibliography{dn}
	
\end{document}